# The Vulnerable Male Brain:

# Men's Spatial Abilities are Condition-Dependent, Sexually Selected Traits

David C. Geary[1], David Giofrè[2], Marcia Collaer[3], Richard A. Lippa[4], &
Lewis G. Halsey[5],

[1]Department of Psychological Sciences, Interdisciplinary Neuroscience, University of Missouri, Columbia, MO, 65211-2500, United States (gearyd@missouri.edu)
[2]DISFOR, University of Genoa, Corso Andrea Podestà, 2, Genoa, Italy
[3]Department of Psychology and Neuroscience, Middlebury College, Middlebury, Vermont USA
[4]Department of Psychology, California State University, Fullerton, Fullerton, CA 92834, USA
[5]School of Life and Health Sciences, University of Roehampton, London SW15 4JD, UK

Correspondence:
David C. Geary
Department of Psychological Sciences
University of Missouri
Columbia, MO 65211-2500
Phone: 573-882-6268
Fax: 573-882-7710
Email: GearyD@Missouri.edu

Acknowledgements: We thank Stian Reimers for his assistance, including the access to the BBC data and Michael Peters for access to mental rotation test data.

**Abstract**

Traits that are exaggerated in one sex relative to the other sex might be more vulnerable to stressor exposure because the development and expression of these traits are costly. Sex differences in such traits should therefore be smaller in populations with high stressor exposure. We tested this prediction in humans, by examining the magnitude of men's advantage in spatial cognition and women's advantage in emotion recognition, across nations that varied in their level of development. As predicted, men's advantage in spatial cognition was larger in nations relatively buffered from stressors. However, in contrast to our prediction, women's advantage in emotion recognition was constant across nations, suggesting aspects of men's cognition might be particularly vulnerable to early or current conditions. The samples were biased toward higher income and healthier individuals for nations in which men's spatial cognition was compromised, thus, we are likely underestimating the effects of living conditions on men's spatial cognition. The results further our understanding of how social and environmental conditions can have sex-specific effects on human cognition.

## 1. **Introduction**

Traits that facilitate intrasexual competition for mates or influence intersexual mate choices are often exaggerated, typically in males but sometimes in females. Darwin (1871), for instance, linked associated sexual dimorphisms in physical size or armaments with male-male competition (Andersson, 1994). Hamilton and Zuk (1982) proposed that sexually selected traits are heritable indictors of fitness or more precisely honest indicators of resistance to deleterious pathogens. In this view, male birds with brightly colored plumage have an enhanced resistance to local parasites. Females who prefer such males benefit through enhanced immunity of their offspring.

The focus of Hamilton and Zuk's (1982) proposal was on genetic variability but variation in sexually selected traits can also reflect variation in social and ecological conditions. Cotton et al.'s (2004) review showed that experimental manipulations of diet quantity and quality, pathogen load, and social competition, among other stressors compromised sexually selected traits more severely than other traits (Geary, 2015; Halsey & Geary, 2025; Zahavi, 1975). A universal mechanism for these stressor effects could be the energy costs of trait development and maintenance based on Hill's (2014) proposal that sexually selected traits are direct signals of mitochondrial functions, including energy production. Where trait exaggeration is apparent in one of the sexes, the key prediction is that the magnitude of sex differences will decrease with stressor exposure especially for traits that are energetically costly to build and maintain (Geary, 2017). If so, sex differences that evolved through natural selection (Lassek & Gaulin, 2022) might also be condition dependent, if the associated traits are energetically costly to build and maintain.

Giofrè et al. (2025) showed that human physical dimorphisms related to intrasexual competition and intersexual choice (e.g., height) are larger in populations that are buffered from stressors. Exaggerated traits are not just morphological; they can also be cognitive (Hone et al., 2020). The assumption here is that elaborated cognitive traits in one sex will require more brain tissue or denser long-range white matter connections between distributed brain regions than the same trait in the opposite sex. These elaborations provide cognitive advantages and result in sex differences but come with the cost of being more vulnerable during their development and will likely function below sex-specific norms under stressful conditions (e.g., illness, nutritional deficits). These costs follow directly from the high metabolic demands of the human brain (Mosharov et al., 2025) - demands that might make sex differences in the brain and cognition especially sensitive to stressors (Geary, 2017).

Sex differences in various forms of spatial cognition provide an illustration. These differences are found in species where one sex navigates the environment in more complex ways than the other, such as to search for mates, as found for meadow voles (*Microtus pennsylvanicus*) (Gaulin & Fitzgerald, 1986; Jašarević et al., 2012; Perdue et al., 2011). In traditional human societies, men navigate larger ranges than women (e.g., as related to hunting) and more frequently engage in other spatial-related activities, such as tool construction and use (Hegarty, 2004; MacDonald & Hewlett, 1999; Miner et al., 2014). These activities could both improve men's spatial cognition and provide ultimate explanations for men's moderate advantages in these domains (Geary, 2021; Nazareth et al., 2019; Voyer et al., 1995).

There appear to be spatial-related brain areas that are more elaborated in men than women, controlling for overall brain size (Koscik et al., 2009; Williams et al., 2021). Williams et al. found that men have larger than expected areas of the occipital and temporal cortex involved in object processing, including understanding their functions as tools (e.g., right fusiform, lateral occipital)( Grill-Spector et al., 2001; Mion et al., 2010), as well as enhancements in areas associated with scene navigation and movement in space (e.g., left retrosplenial cortex and entorhinal cortex)(Foster et al., 2023; Jacobs et al., 2010). In keeping with a potential sex-specific vulnerability, in mice species (e.g., *Peromyscus maniculatus*), where the males are advantaged in spatial cognition, poor prenatal nutrition and toxin exposure compromise males' more than females' spatial learning and memory (Jašarević et al., 2014; Jašarević et al., 2011) and the same might be true for humans (Guo et al., 1995).

Better living conditions also appear to influence the development of spatial abilities, and more so in men than women: Lippa et al. (2010) found that men's advantage in spatial abilities is largest in wealthier nations. However, several factors cloud the interpretation of Lippa et al.'s results. A standard measure of stressor exposure was not used, and participant samples were skewed toward higher socioeconomic groups. In developing nations, wealth inequality is associated with larger differences in men's height across socioeconomic conditions than is found in wealthier nations (Giofrè et al., 2025). As a result, samples that are biased toward more advantaged segments of populations in developing nations will result in smaller cross-national differences in height, underestimating the effects of stressor exposure on trait vulnerability.

Our study provides a broad analysis of socio-environmental correlates of cognitive functions across three traits, two that favor men and one that favors women. We used Lippa et al.'s (2010) British Broadcasting Corporation (BBC) dataset for two tests of spatial cognition, which favor men, and Greenberg et al.'s (2023) dataset on recognizing emotions conveyed through eye expressions, which favor women, to test three hypotheses arising from the theory that cognitive sex differences become exaggerated in more benign environments (Geary, 2015). To assess the environment, we used a standard measure of stressor exposure, that is, the Human Development Index (HDI). Individuals living in low HDI countries are more likely to suffer from infectious diseases, poor nutrition and a larger overall disease burden than are individuals living in high HDI countries (Emadi et al., 2021).

First, we tested the hypothesis that the skew toward higher socioeconomic participants in the BBC dataset would result in an attenuated relationship between height and HDI. This analysis involved comparing cross-national relationships of adult heights and HDI across the BBC dataset and height data from fully representative samples collected by the World Health Organization (WHO, 2002). Sample bias would be confirmed if BBC participants are taller than WHO participants in developing nations.

Second, again using Lippa et al.'s (2010) BBC data, we tested the hypothesis that men's advantage in spatial cognition is larger in populations that are relatively buffered from stressor exposure, as measured by the HDI.

Third, we tested the hypothesis that sex differences in social-cognitive domains that typically favor women are also larger in populations with relatively low stressor exposure: These competencies provide advantages in social contexts, including women's intrasexual relational aggression (Geary, 2021). We used a measure that taps women's advantage in recognizing emotions conveyed through eye expressions and that also engages the brain network related to theory of mind (Greenberg et al., 2023; Schurz et al., 2014). As with spatial abilities, some of the associated areas (e.g., regions of left inferior parietal and inferior frontal cortices) are disproportionately larger in women than men (Bzdok et al., 2016; Williams et al., 2021) but women's advantage might also be related to better network integration (Bos et al., 2016). Whatever the source of their advantage and analogous to men's advantages in spatial cognition, women's advantage in recognizing emotions might make this trait more sensitive to living conditions in women than men.

## 2. Materials

**(a) Data and code availability**

The data were obtained from existing sources. WHO data are from World Health Survey (WHS) microdata and are available from the WHO Multi-Country Studies Data Archive upon application request: https://apps.who.int/healthinfo/systems/surveydata/index.php/catalog/whs. Information on the BBC study also requires a request from data holders: https://www.bbc.co.uk/science/humanbody/sex/articles/spatial_tests.shtml. Data from Reading the Mind in the Eyes Test (Eyes Test) are from Supporting Material (Table S2) from Greenberg et al. (2023), https://www.pnas.org/doi/abs/10.1073/pnas.2022385119.

**(b) Participants**

The BBC data were collected in 2005 and included responses from 255,114 participants. The assessment was a web-based survey of about 200 questions focused on human sex differences and took about 40 min to complete (see Reimers, 2007). Lippa et al. (2010) reported cross-national results for the spatial measures, but they did not examine sex differences using the HDI which has been linked to sex-specific physical changes (e.g., height), and they did not assess the potential influence of sample bias. As in the Lippa et al. study, we restricted the age range to 18 to 60 years and only included countries with at least 30 men and 30 women for every target variable (i.e., mental rotation, line angle judgment, height). This ensured reliable and balanced estimates across sexes within countries. These filtering criteria were defined a priori to reduce the risk of biasing the subsequent analyses

After applying these filters, the final BBC sample included 211,656 participants ($M$ = 31.20 years, $SD$ = 10.27) (45.40% women) (17.04% excluded). A list of the 55 included countries and associated sample sizes is provided in the supplementary material.

The countries represented in the BBC sample were matched to their corresponding HDI for the year 2005. For each country, aggregated statistics were available separately for men and women for performance on the Mental Rotation Task (MRT) and Line Angle Task (LAT), along with height. Sex was coded as male or female based on self-report in the original BBC data.

The Eyes Test scores were obtained from 141,453 women and 135,619 men across 57 nations, ranging from 16 to 70 years of age (Greenberg et al., 2023). These were voluntary participants who competed an English web-based version of the test (https://www.labinthewild.org/). HDI values from 2015 were used to assess cross-national variation in living conditions for the Eyes Test data (most Eyes Test data were collected between 2013 and 2019).

**(c) Measures**

**National Human Development Index.** The HDI is a commonly used measure of national differences in overall quality of life based on the estimated well-being of individual citizens (https://hdr.undp.org/data-center/human-development-index#/indicies/HDI). The key domains are life expectancy at birth, educational opportunities, and overall standard of living (Lind, 2023). The latter is based on gross national income adjusted for purchasing power parity. The prevalence of communicable, maternal, neonatal and nutritional diseases in low (e.g., Sudan), medium (e.g., India), and high (e.g., China) HDI countries is 2.02, 1.80, 1.53 times higher than those found in very high (e.g., United States) HDI countries (Emadi et al., 2021).

**Height.** The height data from the BBC were self-reported in various formats (e.g., feet and inches, m or cm), which were converted to a standard metric scale. To ensure data quality, height was filtered to retain only plausible values; specifically, height entries were transliterated to remove non-ASCII characters, then converted to cm using a custom conversion function. Heights were removed if they fell outside the range of 100 to 250 cm; all other entries were treated as missing. Height data from the WHO were added as a separate variable, obtained from national-level reports and surveys described in Giofrè et al. (2025). The survey produced nationally representative data on adult health and its determinants across 69 countries, using standardized sampling and data collection methods (WHO, 2002), and likely reflect more representative population-level values than do the BBC data. To compare the two height datasets, we retained only those countries present in both the WHO dataset and the BBC sample. Unfortunately, the overlapping samples had a restricted range of HDI scores – very few low HDI countries – relative to Giofrè et al.'s study. Moreover, Giofrè et al.'s analyses included 69 countries from the WHO dataset, but only 35 of these overlapped with the BBC dataset. The smaller sample and restricted range of HDI scores could attenuate the relation between height and HDI, and the sex by height interaction across HDI.

No associated height data were available for the Eyes Test samples and thus sample bias in this dataset could not be systematically explored. The authors suggested there was no substantive bias based on the proportion of participants with worldwide web access relative to national access levels (Greenberg et al., 2023). However, the test was administered in English which could have biased the sample to more educated segments of the population in non-English speaking countries.

**Mental Rotation Task (MRT).** Participants completed a six-item computerized MRT (Peters et al., 1995) like the original Vandenberg and Kuse (1978) test. Each item presented a target three-dimensional figure on the left, along with four response alternatives on the right. Two of the alternatives were correct matches to the target figure, rotated in space, while the other two were distractors. Participants were instructed to identify the two figures that matched the target by mentally rotating them. The total score was the number of correctly identified matching pairs across the six items (max = 12). This task is known to produce robust sex differences in performance, typically greater than $d = .90$ for older adolescents and adults (Masters & Sanders, 1993; Peters et al., 1995). Lippa et al. (2010) reported a more moderate overall (across nations) sex difference ($d = .47$), likely because the task had fewer items than typical MRT measures.

**Line Angle Task (LAT).** Following Collaer et al. (2007), the task assesses visuospatial perception and requires participants to match the angle of the single presented line at the top of the screen to one of 15-line options in an array at the bottom of the screen. The 20 test items were presented one at a time, and the score was the number of correctly matched items. Collaer et al. (2007) found a moderate male advantage for adults ($d = .57$), as did Lippa et al. (2010), $d = .49$.

**Eyes test**. The test is based on Baron-Cohen's (2003) systematizing-empathizing theory of sex differences and includes 36 items that are images of actor's eyes while they are experiencing a specific emotion. Two psychologists generated the target word (i.e., associated emotion) and three foils for each item. Items in which at least five of eight judges agreed on the target emotion were retained for the measure (Baron-Cohen et al., 2001). The task is to identify the corresponding emotion from four options. The score is the number of correctly identified emotions. Across multiple samples and controlling for age, Greenberg et al. (2023) found a small overall advantage for women ($\beta = .17$).

**(c) Quantification and statistical analyses**

The unit of analysis was the mean (height, Eyes Test) or median (BBC cognitive measures) by country for each sex. The full linear model built to investigate each cognitive measure was outcome_variable ~ HDIc + Sex + HDIc*Sex, where HDIc is HDI centered at 0.5 to improve interpretability. To investigate height, the full model also included the additional term 'database source'. The exact sample in each analysis varied depending on data availability for the dependent variable. Visual interpretation of data plots (Loftus 1993) was supported by interpretation of effect sizes and *p* values. The *p* value is highly sensitive to slight differences in a dataset or how it is analysed (Halsey, Curran-Everett et al. 2015), and controlling for long-run error rates by employing a hard cut-off (alpha value) has many limitations (Boos & Stefanski 2011, Halsey 2019). Thus, following Fisher (Fisher 1935) we treat the *p* value as a continuous variable providing an approximate level of evidence against the null hypothesis (Fisher 1959).

All statistical analyses were conducted using R (version 4.5.0; R Core Team, 2025). Linear models were estimated using the lm() function from base R. Visualizations were produced using the 'ggplot2' package (Wickham, 2016).

**3. Results**

Hypothesis 1: The skew toward higher socioeconomic participants in the BBC dataset should result in an attenuated relationship between height and HDI. Indeed, participants in the BBC sample tended to be highly educated. A substantial proportion reported completing secondary school (n = 37,964), a university degree (n = 87,707), a postgraduate or professional degree (n = 33,011), or technical and vocational (n = 23,795) or other college training (n = 26,661), totaling 98.9%. In contrast, relatively few individuals had not completed secondary school (n = 2,364). This suggests an over-representation of individuals from higher socioeconomic backgrounds (see Reimers, 2009), especially for samples from lower HDI countries.

For example, in Egypt in 2005 about 30% of young people continued to tertiary education (www.theglobaleconomy.com) while over 80% of BBC respondents from Egypt reported having done so. In Pakistan, in 2005 less than 10% of young people entered tertiary education (www.theglobaleconomy.com) compared to nearly 90% of BBC respondents from Pakistan. In wealthier countries, the education levels of the BBC respondents were similar to that of their national populations. For example, in the USA in 2005 about 80% of young people reached tertiary education (www.theglobaleconomy.com) and about 80% of BBC respondents reported the same, while in the UK about 60% reached tertiary education in 2005 (www.theglobaleconomy.com) while over 80% of BBC respondents reported the same.

A first linear model was fitted with Sex, Database (BBC vs. WHO), and HDI as predictors of height, including only the main effects. Men were taller than women (β = 11.88, *SE* = 0.43, $p < .001$), participants in the BBC dataset were taller than those in the WHO dataset (β = –38.42, *SE* = 2.70, $p < .001$), and height increased with HDI (β = 22.28, *SE* = 1.82, $p < .001$). The average sex difference in height was 11.88 cm, with a large effect size (Cohen's $d = 3.99$, $\eta^2 = .80$), indicating a robust difference across the sample ($R^2 = .85$). Next, a fully saturated model was fitted, including all possible two- and three-way interactions among Sex, Database, and HDI. While evidence for the main effects remained clear, the only significant interaction was between Database and HDI (β = 32.25, *SE* = 6.08, $p < .001$). The interactions with Sex were not significant (e.g., $p = .73$ for Sex × Database; $p = .79$ for the three-way interaction). A final model was therefore estimated including the main effects and the Database × HDI interaction only. This model explained additional variance in height ($R^2 = .87$), with large effect sizes for Sex ($\eta^2 = .80$), Database ($\eta^2 = .55$), and HDI ($\eta^2 = .47$), and a moderate-to-large effect for the interaction ($\eta^2 = .26$). The significant interaction indicated that the positive association between HDI and height was steeper in the WHO dataset than in the BBC dataset (Figure 1), in keeping with the hypothesis. Every 0.1 unit increase in HDI was associated with an additional 3.34 cm increase in height in the WHO data compared to the BBC data, holding all other variables constant. Thus, the HDI measure likely underestimates stressor exposure in low-HDI countries in the BBC dataset.

Hypothesis 2: For MRT, the first model tested the main effects of HDI and Sex. Performance on the MRT increased as HDI increased ($\beta = 4.65$, $SE = 0.53$, $p < .001$) and men outperformed women ($\beta = 1.50$, $SE = 0.11$, $p < .001$). We then included the interaction between HDI and Sex ($\beta = 2.53$, $SE = 1.03$, $p = .015$) which offered evidence supporting the hypothesis that men gained more than women in MRT performance as HDI increased (figure 2a).

LAT performance increased as HDI increased ($\beta = 3.00$, $SE = 0.58$, $p < .001$) and men performed better than women ($\beta = 1.90$, $SE = 0.12$, $p < .001$). Including the interaction between HDI and Sex ($\beta = 2.15$, $SE = 1.15$, $p = .064$) offered some evidence supporting the hypothesis that men gained more than women in LAT performance as HDI increased (figure 2b).

Hypothesis 3: For the Eyes Test, performance increased as HDI increased ($\beta = 6.22$, $SE = 1.01$, $p < .001$) and women outperformed men ($\beta = 1.00$, $SE = 0.21$, $p < .001$). The interaction term offered no evidence that women gained more on this measure than men as HDI increased ($\beta = 2.03$, $SE = 2.02$, $p = .317$), contra the hypothesis (figure 2c).

**4. Discussion**

As living conditions improve across the world, men gain more than women in height and weight, while women gain more than men in lifespan (Giofrè et al., 2025; Medalia & Chang, 2011). Explanations for larger sex differences in wealthier and healthier countries, which are generally more equalitarian, have focused on a relaxation of social constraints on the expressions of individual preferences (Herlitz et al., 2025; Kanazawa, 2024). A loosening of social constraints is a plausible explanation for some traits, such as personality and occupational preferences (Halsey & Geary, 2025), but it does not explain changing sex differences in height and lifespan; the mechanisms through which weaker gender roles can have sex-specific effects on cognition are also unclear.

The potential for energetically costly and exaggerated traits to be more vulnerable to environmental and social conditions – whether they emerged through sexual or natural selection – might provide a unifying framework for understanding sex differences in physical traits and cognition (Geary, 2015, 2016, 2021; Hill, 2014; Lassek & Gaulin, 2022). One associated prediction is that improved living conditions should result in more gains for men than women in height and weight, as shown previously (Giofrè et al., 2025).

We also analyzed the relations between height and living conditions in the present study to test whether the strength of the overall (across sex) relation between height and HDI was attenuated in the BBC relative to the WHO dataset. We also examined the sex by HDI interaction reported by Giofrè et al. (2025), but this was not significant. The likely reason is that the WHO dataset used here only included half the nations used in the analysis by Giofre et al. and with a truncated range of HDIs. Nonetheless, we found that the relation between HDI and overall height was substantially smaller in the BBC than WHO dataset, which is consistent with sample bias. Just as our analyses indicate that the biased nature of the BBC dataset underestimates relations between HDI and sex differences in height, the BBC dataset might also underestimate the relations between HDI and sex differences in spatial cognition.

As predicted, men's advantage on the spatial measures was larger in higher HDI countries. Contra to our hypothesis, however, women's advantage in recognizing emotions was consistent across countries with varying living conditions. Women typically exceed men in various forms of social cognition, and our results suggest this advantage might be constant across environments. However, severe nutritional stressors, such as those associated with anorexia nervosa, are associated with compromises in women's social cognition, including emotion recognition (Bora & Köse, 2016), and other studies suggest that the magnitude of women's advantage in episodic memory is influenced by living conditions (Asperholm et al., 2019). It might be that females are buffered from moderate stressors at least for emotion recognition, that females are vulnerable to different types of stressors than males, or that there was undetected sample bias in the Greenberg et al. (2023) dataset; a mechanistic understanding is presently lacking.

Whatever the mechanisms, our results suggest that men's spatial cognition might be more sensitive to early or current living conditions than some aspects of women's cognition, at least within the range of living conditions represented in our data. This could be due to poor early conditions compromising prenatal and early postnatal testosterone production in boys (Thompson & Lampl, 2013) which in turn could compromise later spatial abilities. However, prenatal testosterone effects on later spatial abilities are uncertain (Collaer & Hines, 2020). There is, however, some evidence that the postnatal surge in testosterone contributes to sex differences in spatial abilities (Constantinescu et al., 2018), but a consensus has yet to be reached.

Another possibility is that the more rapid male prenatal and early postnatal growth and higher metabolic demands result in steeper trade-offs in the development of the brain networks underlying men's later spatial cognition relative to the networks for emotion recognition (Alur, 2019; Garn et al., 1953; Jašarević et al., 2014). Although face processing and complex spatial processing engage distributed brain networks (Cona & Scarpazza, 2019; Xu et al., 2021), the networks undergirding performance on our spatial tasks might be more costly when engaged than those undergirding performance on the eyes test. Spatial tasks widely engage various parietal regions, and Mosharov et al. (2025) found that some of these spatial areas have a high density of mitochondria, suggesting they are costly to use. The ultimate test will require assessment of the metabolic costs associated with the development and functioning of these networks, which is technically possible (Palombit et al., 2022), but has yet to be done in these domains.

More robust tests of the hypothesis of sex-specific vulnerabilities in cognition would also include assessments of within-nation, cross generational changes in the magnitude of cognitive sex differences. Halsey and Geary (2025) showed that when nations improve the living conditions of their citizens, the population becomes taller and men's advantage in height becomes larger. Improvements in living conditions also result in broad gains in cognition for the population (Pietschnig & Voracek, 2015), and there is some evidence that men gain more in spatial cognition than in other cognitive domains while women gain more in aspects of social cognition (Asperholm et al., 2019; Teasdale & Owen, 2000), but these patterns are only suggestive and require further study.

Finally, it is possible that our results are confounded by women's use of oral or other (e.g., IUD) hormone contraceptives if usage varied across nations. For the BBC data, 21.6% of women reported use of contraceptives (Reimers, 2007) and some studies suggest this could influence performance on spatial tests, but the results are mixed overall (Peters et al., 1995; Wharton et al., 2008). These mixed results are likely related to substantive differences in the route of administration (e.g., patch, oral) and differences in the hormonal formulations of different contraceptives (Beltz, 2024). At this time, it appears that use of contraceptives that include androgenic progestin might enhance women's spatial performance (Yan et al., 2025; but see Jensen et al., 2023), whereas this performance might decline with high estrogen concentrations (Hampson et al., 2022; Peragine et al., 2020). Contraceptive usage was not assessed in the eyes test sample, but the largest study in this area revealed no effect of oral contraceptive usage on eyes test performance and no correlations between this performance and circulating estrogen, progesterone, or testosterone concentrations (Shirazi et al., 2020). In all, any potential confounds due to use of oral contraceptives on spatial cognition cannot be determined from the available data.

In all, we provide evidence for greater effects of environmental stressors on spatial cognition in men, but no evidence for stressors affecting women's recognition of emotions. There is good reason to believe that the relation between these stressors and magnitude of the observed cognitive sex differences, especially for the spatial measures, might be an underestimate of the true vulnerability to stressor exposure, given the likely bias in the BBC and perhaps also the Greenberg et al. (2023) samples. More definitive testing of sex- and trait-specific vulnerabilities will require assessments of specific stressors, such as pathogen load and nutritional deficits, during development and in adulthood. The key message here is that the magnitude of at least some human cognitive sex differences varies in ways consistent with environmental influences on with condition-dependent trait expression (Cotton et al., 2004; Geary, 2015), supporting the potential value of more targeted follow-up studies.

In closing, we note that mean population height and sex differences in height are reliable biomarkers of living conditions (Giofrè et al., 2025; Tanner, 1992), and so future cross-national studies of sex differences in physical, psychological and cognitive traits might include height as a variable and use population height relative to the WHO (2002) dataset to assess for potential sample biases, as in the present study.

**Figures**

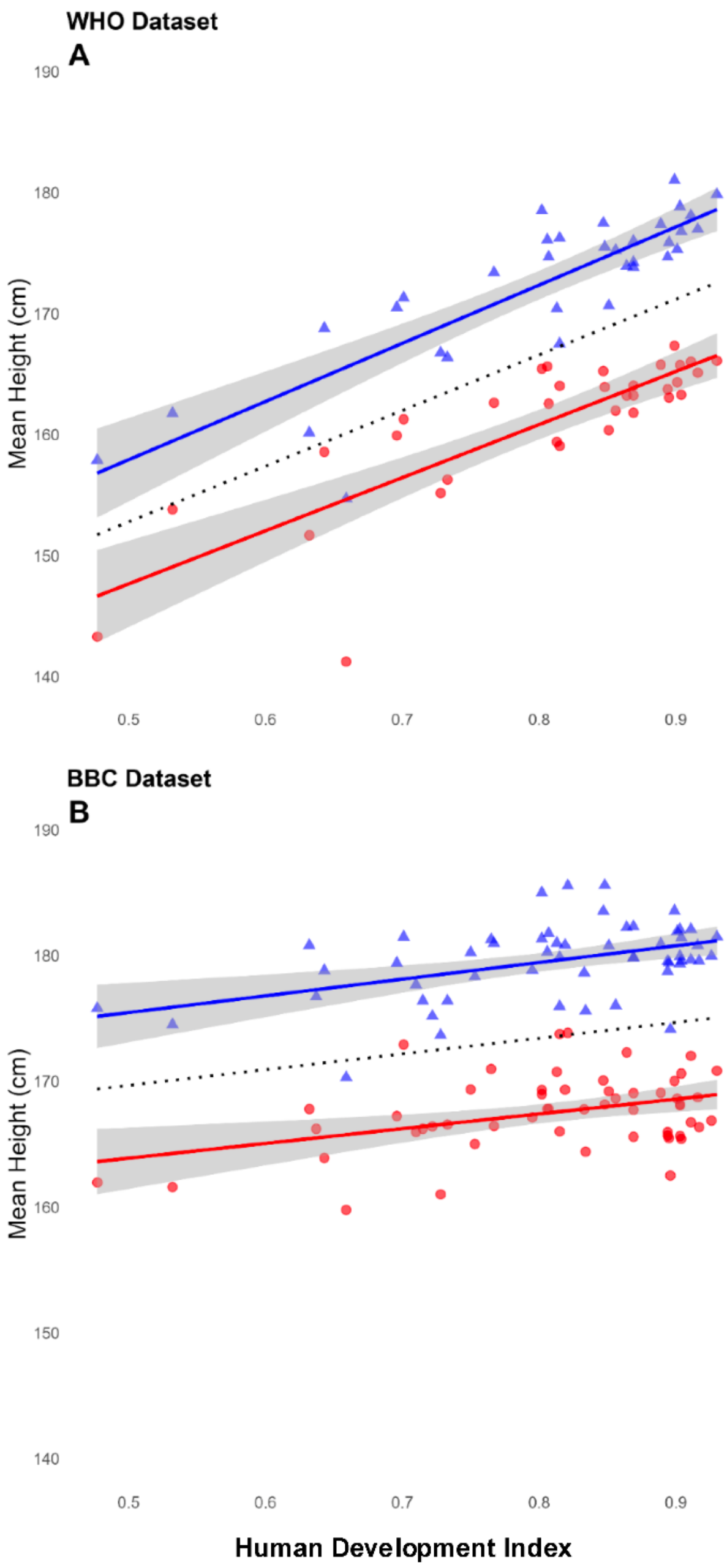

Figure 1. Mean height against human development index (HDI) at the level of individual nations, disaggregated by sex (blue: males; red; females), for two distinct data sets (A: WHO and B: BBC). The colour-coded lines and shading are the best fit regressions and associated 95% confidence intervals. The stippled lines represent the best fit regressions for the two sexes combined. The relationship between height and HDI is steeper in the WHO dataset for both sexes, mainly because height is lower in nations with lower HDIs.

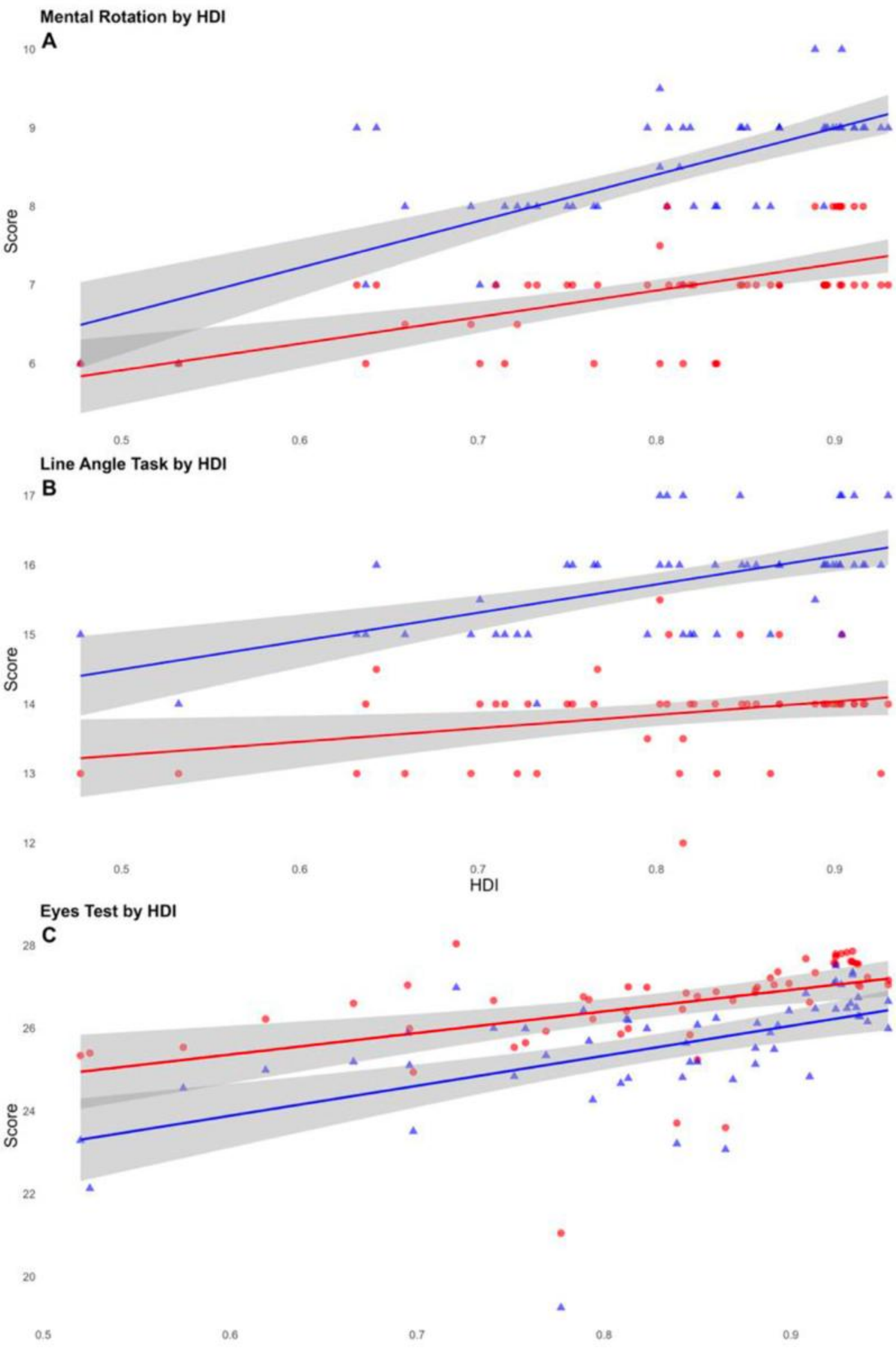

*Figure 2.* A. Mental Rotation Task, B. Line Angle task, C. Eyes Test across HDI, by sex. Shaded areas around the regression lines represent 95% confidence intervals for lines of best fit.